\documentclass{article}

\usepackage{cite}
\usepackage{amsmath,amssymb,amsfonts}
\usepackage{algorithmic}
\usepackage{graphicx}
\usepackage{textcomp}

\usepackage{tikz}
\usepackage{caption,subfig,floatrow}
\usepackage{booktabs,array,multirow}

\usepackage[hidelinks]{hyperref}
\usepackage{cleveref}
\usepackage{comment}

\usepackage{authblk}

\begin{document}

\title{Towards deep learning-powered IVF: 
A large public benchmark for morphokinetic parameter prediction
}

\author[1]{T. Gomez}
\author[2]{M. Feyeux}
\author[2]{J. Boulant}
\author[1]{N. Normand}
\author[2]{L. David}
\author[3]{P. Paul-Gilloteaux}
\author[2]{T. Fréour}
\author[1]{H. Mouchère}

\affil[1]{Nantes University, Centrale Nantes, CNRS, LS2N, F-44000 Nantes, France}
\affil[2]{University of Nantes, Nantes University Hospital, Inserm, CNRS, SFR Santé, Inserm UMS 016, CNRS UMS 3556, F-44000 Nantes, France}
\affil[3]{University of Nantes, Nantes University Hospital, Inserm, CNRS, SFR Santé, Inserm UMS 016, CNRS UMS 3556, F-44000 Nantes, France}
\affil[3]{University of Nantes, Nantes University Hospital, Inserm, CRTI, Inserm UMR 1064, F-44000 Nantes, France}

\maketitle

\begin{abstract}
An important limitation to the development of Artificial Intelligence (AI)-based solutions for In Vitro Fertilization (IVF) is the absence of a public reference benchmark to train and evaluate deep learning (DL) models. In this work, we describe a fully annotated dataset of 704 videos of developing embryos, for a total of 337k images. We applied ResNet, LSTM, and ResNet-3D architectures to our dataset and demonstrate that they overperform algorithmic approaches to automatically annotate stage development phases. Altogether, we propose the first public benchmark that will allow the community to evaluate morphokinetic models. This is the first step towards deep learning-powered IVF. Of note, we propose highly detailed annotations with 16 different development phases, including early cell division phases, but also late cell divisions, phases after morulation, and very early phases, which have never been used before. We postulate that this original approach will help improve the overall performance of deep learning approaches on time-lapse videos of embryo development, ultimately benefiting infertile patients with improved clinical success rates \footnote{Code and data are available at \url{ https://gitlab.univ-nantes.fr/E144069X/bench\_mk\_pred.git}.}
\end{abstract}

\section{Introduction}

Infertility is a global health issue worldwide \cite{infertilityAroundTheGlobe}. The number of couples reporting infertility and referring to assisted reproductive technology (ART) centers for infertility workup and care in Europe is increasing by 8 - 9\% every year \cite{assistedReprodTechInEurope2009}. One of the most common treatments for infertile couples is In Vitro Fertilization (IVF). It consists of controlled ovarian hyperstimulation, followed by ovum pickup, fertilization, and embryo culture for 2-6 days under controlled environmental conditions, leading to intrauterine transfer or freezing of embryos identified as having a good implantation potential by embryologists. The clinical effectiveness of IVF is variable across regions with reported efficiency ranging from 20\% to 40\%. IVF is mainly hampered by the current limitations of embryo quality assessment methods \cite{IVFLowEfficacy}. Indeed, the main embryo quality assessment method is based on morphological evaluation, which consists of daily static observation under the microscope. Although consensus exists for morphological evaluation of embryo development, this method still suffers from a lack of predictive power and inter- and intra-operator variability \cite{highVari1,highVari2,highVari3}.
Time-lapse imaging incubators (TLI) were first released in the IVF market around 2010. They provide continuous monitoring of embryo development, by taking photographs of each embryo at regular intervals throughout its development, ultimately compiling a video giving a dynamic overview of embryonic in vitro development. This technology allows very stable culture conditions and leads to a dynamic annotation of embryonic developmental events, called morphokinetic (MK) parameters, such as, for instance, cell divisions, blastocyst formation, and expansion. Although several studies have reported an association between MK parameters and implantation potential, the clinical usefulness of TLI remains debated \cite{timeLapseCultureWithMorpho,Paulson2018,Armstrong15}. 
Nevertheless, TLI still appears to be the most promising solution to improve embryo quality assessment methods, and subsequently the clinical efficiency of IVF. In particular, the unprecedented high volume of high-quality images produced by TLI systems could be leveraged using modern Artificial Intelligence (AI) methods, like deep learning (DL). 
Indeed, the recent emergence of DL has revolutionized many fields like games \cite{AlphaGo}, computer vision \cite{alexnet}, language processing \cite{attentionIsAllYouNeed}, protein folding \cite{alphafold}, and its advent has set high expectations on its potential for medicine and biology and called for concrete applications. 
Importantly, the question of data sharing is at the center of DL strategies being applied to health data. Indeed, a model can not be reproduced and evaluated externally if the dataset used to train the model is not made available. The main reason behind this common absence of data sharing probably has to do with concerns about data security and maybe to a lesser extent with the made scientific competition. The consequence of a rather “black box” development of DL methods in IVF results in a lack of consensus about which DL architecture to use, with private companies selling and implementing solutions that have not been independently evaluated by the community, raising questions about potential bias and fairness issues for example \cite{interpretableIVF}. Data sharing is therefore of utmost importance to properly implement DL in IVF practice \cite{interpretableIVF}. In this context, we are in dire need of a reference time-lapse dataset and a baseline analysis with the most common DL algorithms, similar to what has been done in other fields \cite{mimic,chexpert,retina,segDataset}.
Several teams have applied DL models in IVF, but with important limitations: either the number of videos was lower than 300 or the number of total images composing the videos was under 150k (\cref{datasetChar}) \cite{WeakSupMorphoKin,predSuccRate,cellCount,BlastCellCount}. 

\begin{table*}[ht!b]
\noindent\makebox[\textwidth]{%
    \begin{tabular}{c|c|c|c|c|c}
    \toprule
    Author&Year&Video nb.&Image nb.&Phases used&Accuracy obtained \\
    \midrule
    Khan et al. &2016&256&150k&1-5 cells&$87\%$ \\
    Moradi Rad et al.&2018&-&224&1-5 cells&$82.4\%$ \\
    Silva-Rodr\'{i}guez et  al.&2019&263&100k&1-5 cel  &$80.9\%$\\
    Kumar Kanakasabapathy et. al.&2019&-&8k&Blasto/No Blasto&$96\%$ \\
    H Ng et al.&2018&-&600k&tStart to t4+&$84.6\%$ \\
    Liu et al, &2019&170&60k&tStart to t4+&$83.8\%$ \\
    Lau et al.&2019&1303&145k&tStart to t4+&$83.65\%$ \\
    \bottomrule
    \end{tabular}
}
    \caption{Dataset characteristics of previous works.}
    \label{datasetChar}
\end{table*}

Additionally, studies used a limited amount of embryonic stages / MK parameters to identify with DL. Finally, and as stated above, the studies did not share their datasets, making their analysis impossible to recapitulate. A shared dataset should be large enough to train powerful deep learning models, contain full videos to make full use of the TLI information, and have highly detailed annotations taking into account a large number of development phases to maximize potential clinical use.

Here we propose a unique reference benchmark that will allow the community to evaluate and compare morphokinetic models and will be a step towards deep learning-powered IVF. 

Our contributions are the following : 
\begin{itemize}
    \item A dataset that contain 704 full videos and a total of 337k images which was sufficient to train and evaluate deep learning models.
    \item Highly detailed annotations with 16 different development phases, from early cell division phases (t2-t5+) as in previous work, but also late cell divisions (t6 to t9+), phases after morulation (tM to tHB), and very early phases (tPNa and tPNf), which, to the best of our knowledge, have never been reported up to now.
    \item Custom evaluation metrics tailored for the morpho-kinetic parameter extraction problem.
    \item Baseline performance using popular simple models like the ResNet, LSTM and ResNet-3D architectures
\end{itemize}


\section{Methods} 

\subsection{Dataset collection}
Between 2011 and 2019, 716 infertile couples underwent Intracytoplasmic Sperm Injection (ICSI) cycles in our University-based IVF center and had all their embryos cultured and monitored up to blastocyst stage with a TLI system. 
To select the videos, we first excluded videos with strictly less than 6 phases annotated to keep only highly detailed videos and then randomly selected $10\%$ of the remaining videos, which constitutes a dataset of 704 videos.
We subsequently extracted all focal planes using an Application Programming Interface (API) provided by the TLI manufacturer (Vitrolife©). We acknowledge that only ICSI cycles were included in our time-lapse devices over that period, as we considered that conventional IVF would lead to different developmental timings as compared to ICSI. We do not routinely use assisted hatching. There were no major lab changes over the study period. The Local Institutional Review Board (GNEDS) approved this project. All patients agreed with the anonymous use of their clinical data.
Patient treatment and embryo culture protocol were described in a previous study \cite{Freour2015}. In brief, embryo culture was performed from fertilization (day 1) up to blastocyst stage (day 5 or day 6) at 37◦C with 5\% O2 and 6\% CO2 in a sequential culture medium, i.e. G1 plus (Vitrolife©, Sweden) from day 0 to day 3, followed by G2 plus (Vitrolife©, Sweden). We acknowledge that culture media might impact embryo development and have an evolving composition throughout embryo development. However, the available literature does not support the concept of medium-dependent morphokinetic patterns \cite{basile2013type}. Although we agree that there is a need to clarify IVF culture media composition to enhance our understanding of embryo development \cite{sundeSerious2016}, there is no evidence to our knowledge that the content of commercial culture media changes over time in ways that are important enough to consider.
The images were acquired with a TLI system (Embryoscope©, Vitrolife©, Sweden) every 10 to 20 min by a camera under a 635 nm LED light source passing through Hoffman’s contrast modulation optics.
The information about embryo viability is not used in this work as the purpose is to focus solely on morphokinetic parameter prediction. These discarded embryos allowed us to study a variety of abnormal embryonic features (abnormal morphology, abnormal fertilization/number of pro-nuclei, necrosis, fragmentation, developmental delay, etc.) or problems during image acquisition (sharpness, change of focus, brightness, etc.). 
Although we included all available focal planes in the dataset, we only used the center focal plane in our experiments, as we aim to propose baseline models and leave the exploitation of the other focal plans to future research.

\subsection{Dataset annotation}
Each video was annotated by a qualified and experienced embryologist undergoing regular internal quality control. 
For each video, the annotation consists of the timing of 16 cellular events noted tPB2, tPNa, tPNf, t2, t3, t4, t5, t6, t7, t8, t9+, tM, tSB, tB, tEB and finally tHB.
We use the definition of the events proposed by Ciray et al. \cite{ciray} : polar body appearance (tPB2), pronuclei appearance and disappearance (tPNa and tPNf), blastomere division from 2-cell stage to >8 cells-stage (t2,t3,t4,t5,t6,t8 and t9+), compaction (tM), blastocyst formation (tSB, tB), expansion and hatching (tEB and tHB). 
We chose to use more events than previous work \cite{WeakSupMorphoKin,predSuccRate,cellCount,BlastCellCount} to develop models that can more precisely describe the embryo development in the controlled environment. 
We started prospective annotation of the database according to this reference work in 2014, while annotations made before 2014 were retrospectively checked. 

\subsection{From event timing to frame labels}

We formulate the task as an image classification problem.
This means that we need to assign a label to each frame that the model will be trained to predict.
However, the annotation given by the biologists are timings in hours post fertilization that indicates the temporal position of events in the video.

\begin{figure*}
    \centering
    \includegraphics[width=\textwidth]{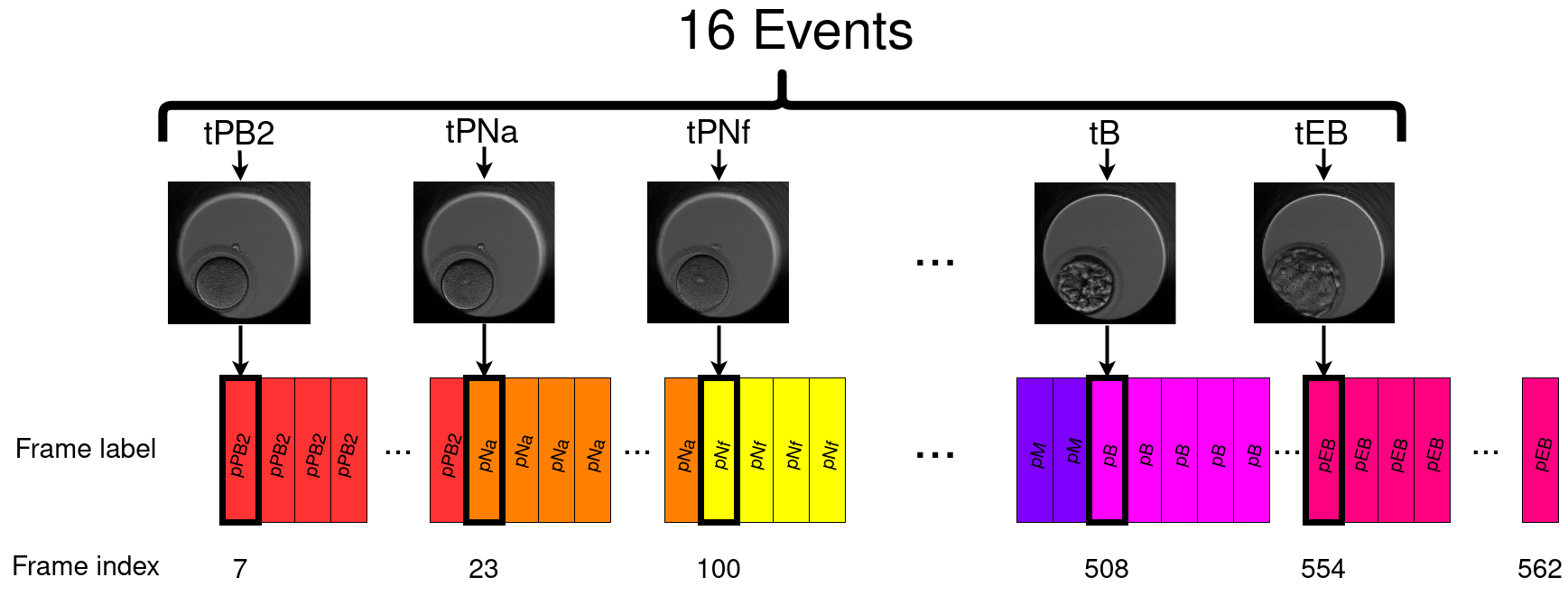}
    \caption{The method used to assign a label to every frame of the video. 
    First we identify at which frame each event occurs and assign to these frame a label corresponding to the event they show.
    The other frames are assigned the label corresponding to the most recent event that has occurred in the previous frames.
    }
    \label{from_timing_to_labels}
\end{figure*}

Knowing the timing at which each frame was taken, we identify the frames corresponding to each event and assigns them a label corresponding to the event they show (noted pPB2, pPNa, pPNf, p2, p3, p4, p5, p6, p7, p8, p9+, pM, pSB, pB, pEB or pHB), as illustrated in \cref{from_timing_to_labels}.
The other frames are assigned the label corresponding to the most recent event that has occurred in the previous frames.

\subsection{From frame labels to event timing}

Once a model has done an inference on all frames of a video, we have a sequence of outputs, where each output is a distribution over the possible labels.
To compare the model predictions to the ground-truth timings, we need to convert the sequence of outputs into a list of timings.
Simply selecting the phase with the maximum score at each frame is not a good solution as it can produce sequences of events that are biologically impossible. 
Indeed, the models used do not have a strong constraint forcing them to respect the chronology of embryo development phases and this can lead to backward transitions (for example $p3 \rightarrow p2$, $pM \rightarrow p9+$, etc.). 
Therefore, we propose to use the Viterbi algorithm to solve this problem, as is often done in the literature \cite{WeakSupMorphoKin,cellCount,BlastCellCount,ICLRMorpho}.
The algorithm makes the prediction consistent by combining the sequence of probabilities produced by the model with a $16\times16$ probability matrix indicating at row i and column j the probability to transition from label i to j at the next frame. 
This matrix is computed empirically on the training set of the model.
Given that biologically impossible sequences of events never occur in the training set, their probability is set to zero in the Viterbi algorithm never predicts such sequences.

The Viterbi algorithm produces biologically plausible sequences of prediction such that all the frames assigned to a label are comprised in an interval only containing frames assigned with this label.
Finally, we construct a list of timings by extracting the timing of the first frame assigned with each label.
Note that a model can sometimes miss an event or predict an event that did not happen.
We explain in \cref{metrics} how we handle these cases.

\subsection{Baseline models}

\begin{figure}[ht!b]
    \centering
    \subfloat[The ResNet model\label{res18}]{
        \centering
        \includegraphics[width=0.65\textwidth]{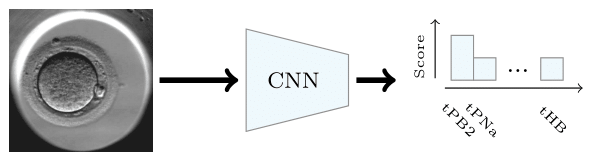}
    }\\
    \subfloat[The ResNet-3D model\label{res3D}]{
        \centering
        \includegraphics[width=0.65\textwidth]{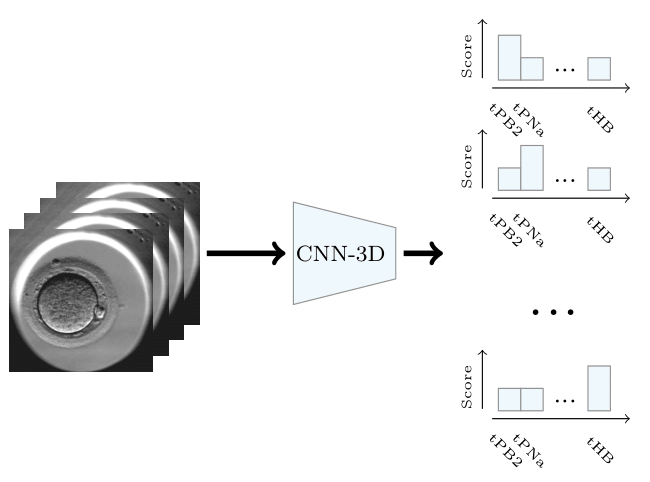}
    }\\
    \subfloat[The ResNet-LSTM model\label{lstm}]{
        \centering
        \smallskip
        \includegraphics[width=0.65\textwidth]{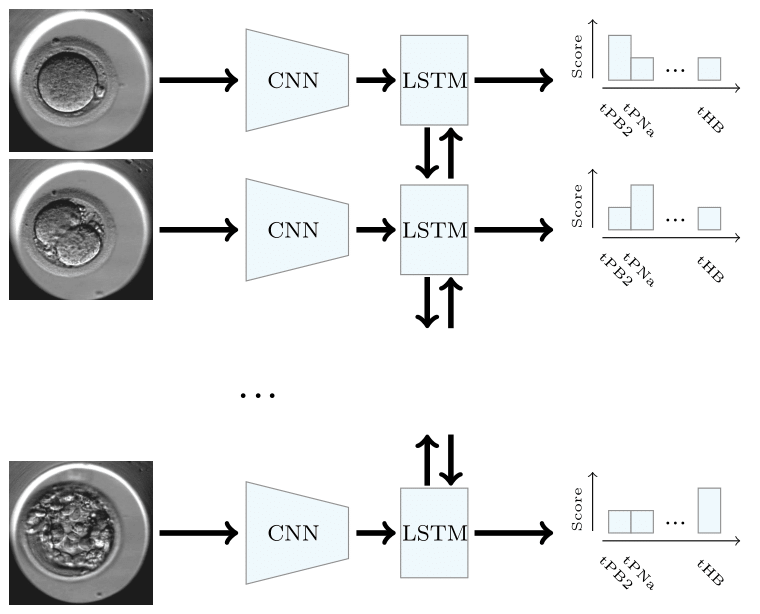}
    }
    \caption{The different models evaluated. ResNet takes an isolated image as input and outputs a vector of class probabilities. ResNet-LSTM and ResNet-3D have as input a short sequence of images and output a sequence of probability vectors.\label{models}}
\end{figure}

Several baseline models were used to perform this classification task and compared using the defined metrics on the annotated dataset. The first model is designed for isolated image classification, the next two models allow the classification of images in a sequence. They are illustrated in \cref{models} and detailed below.

The ResNet Model. Residual models are widely used for the classification of isolated images, for example on ImageNet \cite{resnet}. This model is composed exclusively of convolution layers and contains residual connections every 2 layers. The resolution and the number of channels of the feature maps are respectively divided and multiplied by 2 every 4 layers. After the convolutions, an average-pooling layer produces a vector of features, to which the final soft-max layer is applied to make predictions. We use the variant ResNet proposed by He et. al. \cite{resnet}.

The ResNet-LSTM. This model is the combination of the ResNet model with an LSTM \cite{lstm}. The LSTM model has been designed to model sequences and has been successfully applied in tasks such as speech recognition \cite{LSTMOnVoice}. Pre-activations of the penultimate layer of ResNet are used as a feature vector and are passed to a bi-directional two-layer LSTM that models the evolution through time steps. The size of each hidden unit is 1024. A linear layer after the LSTM calculates the class scores for each image.

The ResNet-3D \cite{resnet3D} is a variant of ResNet designed for the classification of image sequences. This model processes the image sequence by merging temporal information at all layers in the network, allowing both late and early merging of information. For this application, the max-pooling and stride parameters are set to 1 in the temporal dimension. The removal of temporal aggregation is necessary to obtain one prediction per frame in the input sequence. We use the variant ‘R2plus1d-18’ proposed by Hara et al. \cite{resnet3D}.

\subsection{Custom metrics \label{metrics}}

Along with the data and the annotations, we propose several metrics to evaluate models on this benchmarks. 
The first metric is the linear correlation $r$ between the predicted transition timings and the actual timing of the corresponding transitions. 
Before being computed, it requires applying the Viterbi algorithm beforehand so that predictions of the models are made consistent throughout the video. 
The correlation $r$ is computed as follows :

\begin{equation}
r = \frac{C}{V_p\times V_{gt}},
\end{equation}

, where $C$, $V_p$, and $V_{gt}$ are respectively the covariance between the predicted and actual transition times, the variance of the predicted times, and the variance of the actual times. For this metric, only the transitions present in both ground-truth and predictions are taken into account. 
We observed in our experiments that this metric is positively biased, which led us to introduce three more metrics: the accuracy p, the Viterbi accuracy, and the temporal accuracy.
The accuracy p is one of the most widely used metrics in image classification and is defined by the proportion of images correctly labeled by the model:
\begin{equation}
p = \frac{N}{N_{total}},
\end{equation}

where $N$ and $N_{total}$ are respectively the numbers of images correctly classified and the total number of images.
We also define a variant, the Viterbi accuracy $p_v$ that consists to compute the accuracy once the Viterbi algorithm has been applied :
\begin{equation}
p_v= \frac{N_v}{N_{total}} 
\end{equation}

where $N_v$ is the number of correctly classified images once the raw predictions have been refined using the Viterbi algorithm.

Finally, we define the temporal accuracy pt as the average proportion of phase transitions that are predicted sufficiently close to the corresponding actual transition. By “close enough”, we mean that the time separating the predicted transition timing and the actual transition timing is inferior to a threshold. 
This metric also requires that the predictions are made consistent using the Viterbi algorithm and is computed as follows:
\begin{equation}
p_t=\frac{T-T_{far}}{T},    
\end{equation}

where $T$ is the total number of phase transitions and $T_{far}$ is the number of transitions predicted temporally too far away in time from their actual timing.

For example, consider a video containing T = 6 transitions (p2 $\rightarrow$ p3 $\rightarrow$ p4 $\rightarrow$ p5 $\rightarrow$ p6 $\rightarrow$ p7 $\rightarrow$ p8) where the model has predicted the sequence (p2 $\rightarrow$ p3 $\rightarrow$ p4 $\rightarrow$ p5 $\rightarrow$ p6 $\rightarrow$ p8). The model has skipped phase p7 but has predicted phase p8 and this is likely due to the length of phase p7 which can be very short. As, in reality, the embryo cannot skip phase (the fact that some phases cannot be seen in the video is due to the large time interval between successive images), we consider that the model has implicitly predicted p7 with the same timestamp as the one predicted for p8. 
Now, let’s say the transitions (p2 $\rightarrow$ p3) and (p3 $\rightarrow$ p4) are predicted too far away from the corresponding transitions, i.e. the first image where the model has assigned the label of the new phase and the actual image corresponding to the new phase are separated by a time interval superior to a threshold $\theta$. Then, we have $T_{far}=2$ and the temporal precision is $p_t=(6-2)/6 = 0.67$.

The threshold $\theta$ needs to be dependent on the phase because some phases are more difficult to locate precisely in time than others. For this, we use intra-operator standard deviations extracted from Martínez-Granados et al. to have thresholds that are more or less large according to the intrinsic ambiguity of each phase \cite{stdsThres}. In this work, the authors have sent time-lapse videos of embryo development to several IVF centers as an external quality control program and notably studied the intra-operator variance. Using their supplementary data, we compute the standard deviation $\sigma_p$ observed between operators for each phase $p$. The threshold $\theta_p$ we use for phase $p$ is simply set to $\sigma_p$.

The standard deviations for each phase are available in \cref{stds}. The $p$ and $p_v$ metrics have the disadvantage of penalizing models that offer phase transitions far from true transitions as much as those that are wrong by only a few frames. The temporal accuracy metric takes this into account: a model that predicts a phase change close to the actual phase change is favored over a model that is far from the truth.

\begin{table}[ht!b]
    \centering
    \begin{tabular}{*1c|*{7}c}\toprule
        Phase     & pPNa& pPNf& p2   & $p3$& $p4$& $p5$& $p6$\\
        $\sigma_p$& $1.13$    &$0.50$&$0.91$ &$1.81$ &$1.34$ &$1.49$&$1.61$\\
        \midrule
        Phase & $p7$& $p8$& $p9+$& $pM$& $pSB$& $pB$ & $pEB$ \\
        $\sigma_p$ &$2.93$ &$5.36$ &$4.42$ &$5.46$ &$3.78$ &$3.29$ &$4.85$ \\
        \bottomrule
    \end{tabular}
    \caption{Inter-operator standard deviation of annotations in hours. Computed using data from \cite{stdsThres}.}
    \label{stds}
\end{table}

\subsection{Experimental setup}

To show the potential use of this dataset we trained several deep neural networks on our dataset and evaluated them using cross-validation (k = 8). Details of the experiments are given below.

\paragraph{Pre-processing.} The pre-processing procedure we use is largely inspired from the standard procedure proposed in \cite{alexnet}.
During training images are resized from $500 \times 500$ to $250 \times 250$ to reduce GPU memory usage, then a random crop of size $224\times224$ is extracted and finally the images are flipped vertically with a 0.5 probability and flipped horizontally with a 0.5 probability.
During validation and test, the images are also resized to $250\times250$, followed by a center crop of size $224\times224$.
\paragraph{Video selection.} Some embryos grow slowly and the recording of the video only lasts a fixed amount of time, making the early phases over-represented. To prevent the model from being too biased towards early phases, we kept only the videos showing at least 6 
stages of development. The final number of videos used in the experiments was 704.

\paragraph{Hardware and software.} We use two T4 GPUs with PyTorch version 1.10.0.

\paragraph{Hyper-parameters.} Each training batch is composed of 10 sequences of 4 consecutive images. The ResNet model processes each image independently and therefore reads the $10\times4 = 40$ images as if they were independent images. An equal input sequence number and equal sequence length for the three models allow a fair comparison. 
We use a batch size of 10 as we observed during our experiments that it is low enough to train on sequences of images and high enough to provide reasonable training convergence.
The sequence length is set to 4 as it is the maximum length possible given the GPU memory available to us.
The position of the sequence in the video is chosen randomly in the video. We use the standard cross entropy function, optimized with SGD, with a constant learning rate and a momentum. 
The learning rate and momentum values were set to the standard default values proposed by PyTorch.
We applied dropout \cite{dropout} on the last layer of each model during training with a probability $p = 0.50$ which is also the default Pytorch value.
During test and validation, to reduce GPU memory usage, the evaluation batch size was set to 150. The models were not evaluated over the entire video at once but over 150-frame sequences. Since each video contains on average about 500 frames, a few inferences are sufficient to analyze an entire video. Let N be the total number of training frames and L be the number of frames in a sequence. An epoch ends when the model has seen N/L sequences. To select the sequences, we used uniform random sampling with replacement, i.e. the model may see the same image several times and may not see some images within an epoch.
For each split, we used 664, 47, and 45 videos for training, validation, and test. This represents respectively 297k, 20k, and 20k images and allows to detect an absolute variation of 1.5\% during evaluation, with a base accuracy ranging from 65\% to 70\%, a significance level of 0.05, and a power of 0.9, according to a power test.
A model is trained during 10 epochs. The best model considering the validation set is then restored and evaluated on the test set.

\paragraph{Initializing weights.} The ResNet and ResNet-3D weights are pre-trained on ImageNet \cite{alexnet} and Kinetics \cite{resnet3D} respectively. The weights of the ResNet component of ResNet-LSTM are also pre-trained on ImageNet and the weights of the LSTM components are initialized randomly.

\section{Results} 

\paragraph{Compiling a fully annotated, open dataset}. To fill an important gap in the implementation of deep learning in IVF, we sought to build an open resource of fully annotated images of human preimplantation development. The dataset contains 704 videos with annotations for 16 morphokinetic events, covering the whole development of the embryo from day 1 to day 5-6 (Figure 2). This is accompanied by 4 custom evaluation metrics and 3 baseline model performances, along with cross-validation splits to reproduce our results and to rigorously evaluate new models and methods.

\begin{figure*}[ht!b]
    \centering
    \includegraphics[width=0.8\textwidth]{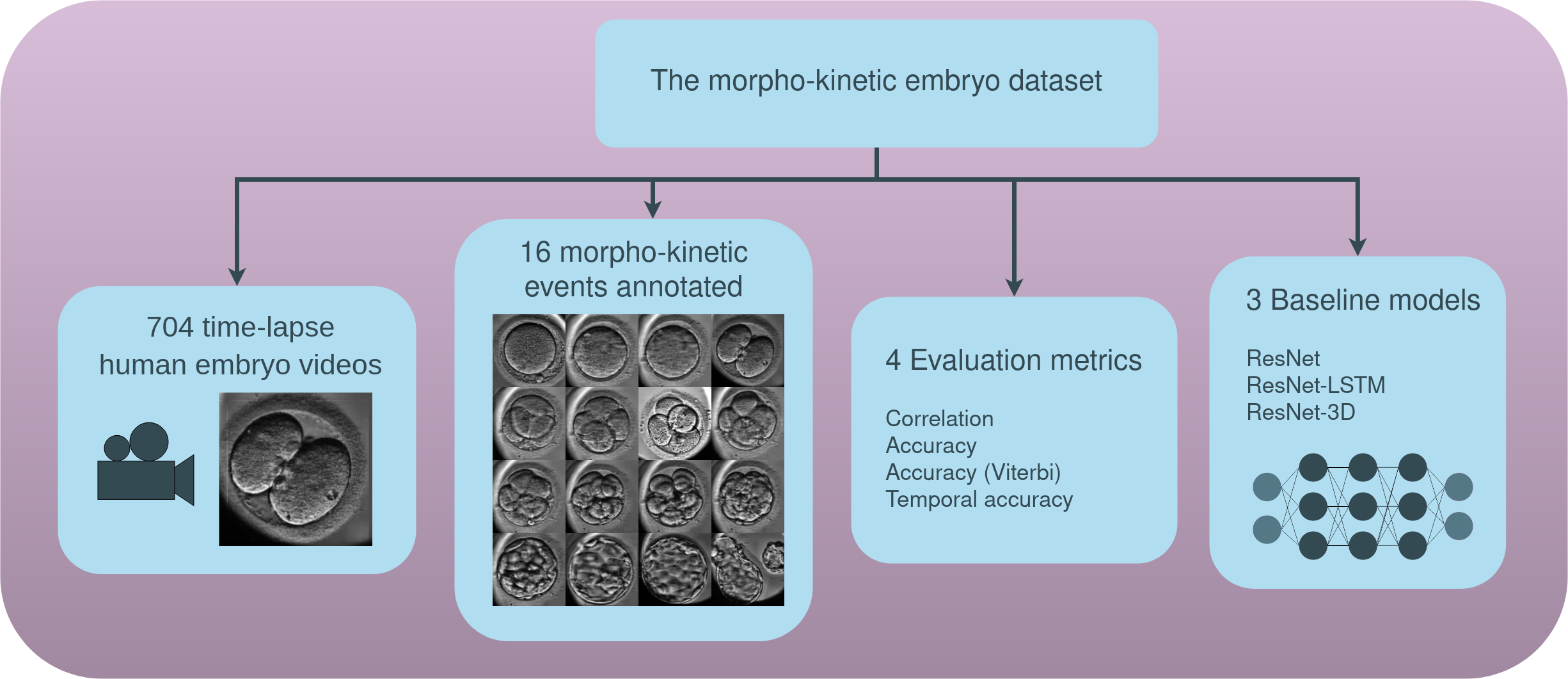}
    \caption{The time-lapse embryo dataset. This dataset contains 704 videos with annotations for 16 morpho-kinetic events, accompanied by 4 custom evaluation metrics and 3 baseline model performances, along with cross-validation splits.}
    \label{workflow}
\end{figure*}

\begin{figure}[ht!b]
    \subfloat[]{\includegraphics[width=0.45\textwidth]{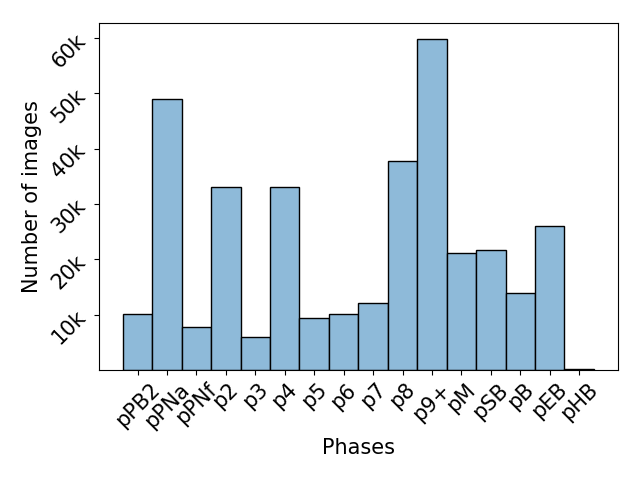}}
    \subfloat[]{\includegraphics[width=0.45\textwidth]{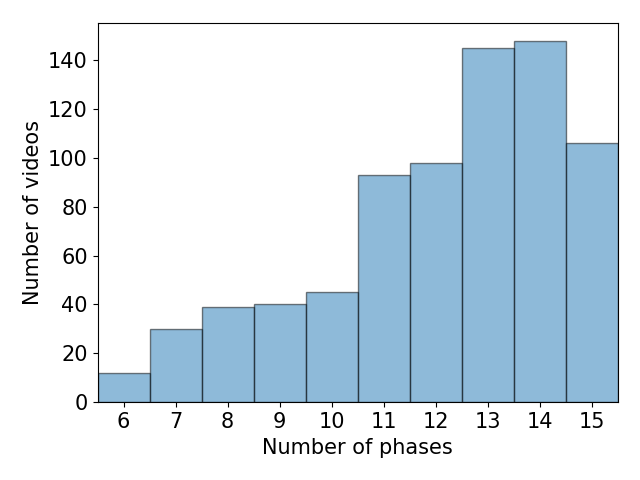}} 
    \caption{Statistics of the dataset. (a) The number of images per phase in the dataset. (b) The number of phases per video in the dataset.}
    \label{fig:classStat}
\end{figure}

Deep learning models are heavily dependent on data and might provide poor performance on a specific class if the amount of input corresponding to it is too small. This is why for each phase, we provide at least several thousand images, even for short phases like pNf, p3, or p5, (\cref{fig:classStat} (a)). The only exception is pHB as it is difficult to capture, the time-lapse recording being often interrupted before reaching that stage. Nevertheless, we still provide more than a hundred images for this phase.
Most videos have at least 8 annotated phases and that approximately 380 videos have more than 13 phases annotated, illustrating the richness of annotation of our dataset (\cref{fig:classStat} (b)). 

\begin{figure*}[ht!b]
\makebox[\textwidth]{
\begin{tabular}{cccccccc}
\includegraphics[width=0.110\textwidth,trim={0.5cm 0.5cm 0.5cm 0.5cm},clip]{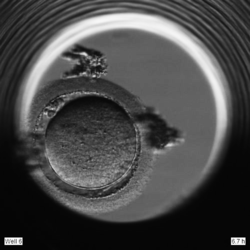} & 
\includegraphics[width=0.110\textwidth,trim={0.5cm 0.5cm 0.5cm 0.5cm},clip]{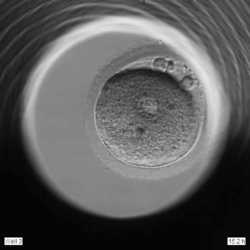} & 
\includegraphics[width=0.110\textwidth,trim={0.5cm 0.5cm 0.5cm 0.5cm},clip]{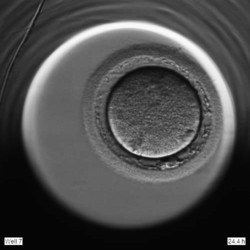} & 
\includegraphics[width=0.110\textwidth,trim={0.5cm 0.5cm 0.5cm 0.5cm},clip]{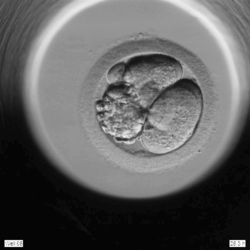} &
\includegraphics[width=0.110\textwidth,trim={0.5cm 0.5cm 0.5cm 0.5cm},clip]{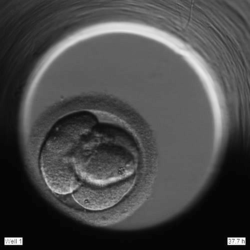} &
\includegraphics[width=0.110\textwidth,trim={0.5cm 0.5cm 0.5cm 0.5cm},clip]{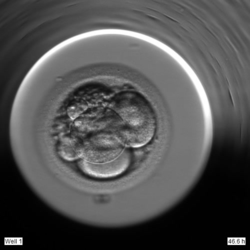} &
\includegraphics[width=0.110\textwidth,trim={0.5cm 0.5cm 0.5cm 0.5cm},clip]{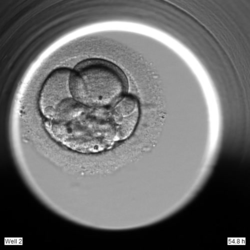} &
\includegraphics[width=0.110\textwidth,trim={0.5cm 0.5cm 0.5cm 0.5cm},clip]{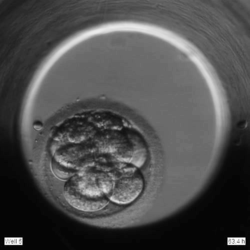} \\

pPB2 & pPNa & pPNf & p2 & p3 & p4 & p5 & p6 \\
Second polar & Pro-nuclei & Pro-nuclei & \multirow{2}{*}{$2$ cells} &\multirow{2}{*}{$3$ cells} & \multirow{2}{*}{$4$ cells} & \multirow{2}{*}{$5$ cells} & \multirow{2}{*}{$6$ cells} \\
body detached  & appearance & disappearance & \\
\includegraphics[width=0.110\textwidth,trim={0.5cm 0.5cm 0.5cm 0.5cm},clip]{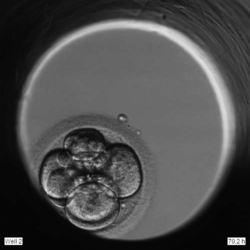} &
\includegraphics[width=0.110\textwidth,trim={0.5cm 0.5cm 0.5cm 0.5cm},clip]{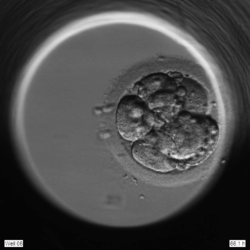} &
\includegraphics[width=0.110\textwidth,trim={0.5cm 0.5cm 0.5cm 0.5cm},clip]{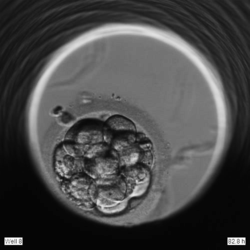} &
\includegraphics[width=0.110\textwidth,trim={0.5cm 0.5cm 0.5cm 0.5cm},clip]{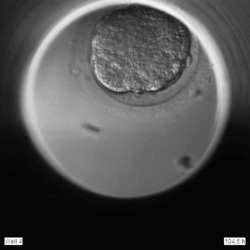} &
\includegraphics[width=0.110\textwidth,trim={0.5cm 0.5cm 0.5cm 0.5cm},clip]{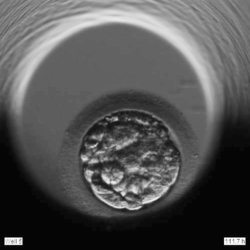} &
\includegraphics[width=0.110\textwidth,trim={0.5cm 0.5cm 0.5cm 0.5cm},clip]{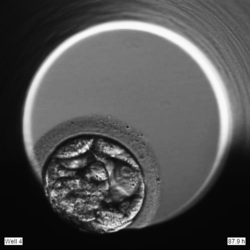} &
\includegraphics[width=0.110\textwidth,trim={0.5cm 0.5cm 0.5cm 0.5cm},clip]{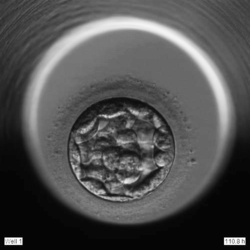} &
\includegraphics[width=0.110\textwidth,trim={0.5cm 0.5cm 0.5cm 0.5cm},clip]{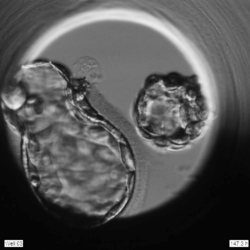} \\
p7 & p8 & p9+ & pM & pSB & pB & pEB & pHB  \\ 
 \multirow{2}{*}{$7$ cells} & \multirow{2}{*}{$8$ cells} & $9$ cells & End of &   Start of   & Full       & Expanded   & Hatching  \\
& & or more   & compaction &  blastulation& blastocyst & blastocyst & blastocyst  \\
\end{tabular}
}

\caption{Illustrations of the $16$ development phases used. Contrast and luminosity are standardized for better visualization.\label{fig:phases}}
    
\end{figure*}

Sample images allow one to have a clear view of the content of the dataset and the annotations associated with the images (\cref{fig:phases}). Note that, depending on their position in the well, embryos can sometimes be partially occluded which is quite common in time-lapse videos. However, even when a part of the embryo is hidden, the images are sufficient to identify the development phase. The behavior of the AI-based model on a specific type of input (like partial views/artifacts) is conditioned on its presence/absence in the training set. In our study, such events were included in the training set, the model was therefore theoretically not affected by these outlier images.

Using the API provided by Vitrolife, we could extract full-length videos and all focal planes available, highlighting the importance of data accessibility. The expert annotations recapitulate embryo development from tPB2 to tHB instead of focusing solely on early cleavages (t2 to t5+), as is done usually in the literature. Finally, 526 videos of our dataset correspond to embryos that were evaluated as compatible with clinical use, i.e. transferred or frozen, which are accompanied by a detailed outcome annotation (results of HCG test, presence/absence of fetal heartbeat, gestational sacs, and live-born information). This means our dataset can also be used by researchers to evaluate outcome prediction models and test cross-center generalizability. Note that the outcome information is considered as patient information, and therefore requires an agreement (MTA) with the University-Hospital of Nantes (see authors for details).

\paragraph{Baseline performance.} The first task to implement deep-learning on a dataset is to train baseline models with the most popular deep-learning architectures. The metrics associated with ResNet, ResNet-LSTM, and ResNet-3D analysis of our dataset are compiled in \cref{full_res}. 

\begin{table*}[ht!b]
\makebox[\textwidth][c]{
\begin{tabular}{cccccc}
\toprule
Model& Image processing & $r$&$p$&$p_v$&$p_t$\\ 
\midrule 
ResNet & Isolated &  $0.961\pm0.026$&$0.663\pm0.041$&$0.701\pm0.044$&$0.371\pm0.09$\\ 
ResNet-LSTM & As a sequence & $\mathbf{0.977\pm0.009}$&$0.685\pm0.041$&$0.696\pm0.043$&$0.559\pm0.223$\\ 
ResNet-3D & As a sequence & $0.97\pm0.021$&$\mathbf{0.705\pm0.036}$&$\mathbf{0.735\pm0.042}$&$\mathbf{0.659\pm0.154}$\\ 
\bottomrule
\end{tabular}}
\caption{Performance of the baseline models obtained after the 5-fold cross-validation. $r$ is correlation, $p$ is accuracy, $p_v$ is Viterbi accuracy and $p_t$ is temporal precision.\label{full_res}} 
\end{table*}

The first metric we considered is the correlation metric, which is close to 1 for all deep-learning approaches. This metric is poorly informative. Indeed, one can notice the high bias and low variance of the correlation metric r that is pushing all values close to 1. This is because the predicted transitions are forced to be in a biologically plausible order after applying the Viterbi algorithm, implying a minimum level of alignment with the actual transitions, hence the high correlation values.
To get a better idea of the performance of the deep learning algorithms,  we focused our analysis on other metrics. The accuracy provides a greater range of values and shows that ResNet, ResNet-LSTM, and ResNet-3D are respectively able to correctly classify on average 66.3\%, 68.5\%, and 70.5\% of the images of the test videos. Logically, the accuracy with Viterbi $p_v$ yields higher values than the regular accuracy p because the model’s predictions are first made biologically plausible.
Finally, the temporal precision shows that ResNet, ResNet-LSTM, and ResNet-3D predict respectively 37.1\%, 55.9\%, and 65.9\% of the transitions at a timing close from the real one.
These 3 metrics highlight the superiority of ResNet-LSTM and ResNet-3D over ResNet. This is explained by the fact that ResNet processes images in an isolated manner, like an embryologist having a static view of the embryo using a microscope, whereas ResNet-LSTM and ResNet-3D process several images together, like an embryologist using a TLI system, and are therefore able to more accurately understand at which development phase the embryo is currently at.
Given that the baseline models ResNet-LSTM and ResNet-3D designed for video processing overperform the model designed for image processing ResNet, this highlights the relevance of proposing a dataset composed of full videos instead of isolated images.
Also, the baseline models achieved good performance which demonstrates that our dataset is sufficient in size and quality to train and evaluate deep learning models. 

\section{Discussion} 

In this study, we propose a large dataset of time-lapse videos of embryo development and make it publicly available for the sake of facilitated and improved further research in the field.
This dataset is accompanied by detailed morpho-kinetic annotations and custom metrics.
We also that simple baseline models can be trained to good performance, showing that the dataset is large enough to train a deep learning model. 
Also, by leveraging image sequence models like ResNet-3D or ResNet-LSTM, we could improve prediction quality, demonstrating the relevance of proposing full videos instead of isolated images.

The good performance can seem surprising as this dataset is composed of only 704 videos and video classification can be considered more data demanding than image classification. However, video classification consists in passing a sequence of images to a model and training it to produce a single output where each video has a single label for all its images. Here, the models are also passed as a sequence of images but they are trained to produce one output per image, i.e. classify each image, and each image has its own label. This is why this problem can be considered as image classification. This dataset size (337k images) is consistent with the dataset size found in the literature, where the number of images ranges from 60k to 600k \cite{WeakSupMorphoKin,predSuccRate,cellCount,BlastCellCount,ICLRMorpho,DLDP}. Kanakasabapathy et al. reported that inter and intra-variance were too high when more than 6 embryo developmental phases were used \cite{inexpAutoDeepLearn}. Contrarily, we report here for the first time the analysis of videos consisting of 16 precise developmental phases. Although some variance is also found in our work, we were able to precisely reconstitute the succession of all 16 morpho-kinetic events. Our work, therefore, goes beyond the simplified set of classes previously used, for example taking only into account early phases up to p9+ and merging phases p4 to p9+ into one class p4+. 
We also show in appendix that this dataset is similar in difficulty to previous datasets by re-evaluating the baseline models using sets of classes usually used in previous work.
Another interesting part of our work is that we implemented 2 improvements to the DL approach. Firstly, we performed 8 fold cross-validation, while previous studies used a single split. Secondly, we used both 3D CNN architecture and a dedicated temporal model, which are relevant considering the temporal nature of the data, leading to improved performance. Although not evaluated up to now in the field of IVF and time-lapse videos of embryo development, the ResNet-3D architecture has been successfully used in several other medical domains such as oncology \cite{Yuan2020,10.1007/978-3-030-33676-9_26} cardiology \cite{10.1117/1.JBO.25.9.095003}, Computerized Tomography (CT) imagery quality, \cite{choi2019multidimensional}  and neuroimaging \cite{10.1007/978-3-030-00689-1_9,10.1117/12.2549758}.
Moreover, 526 of the 704 videos proposed correspond to transferred embryos and can be accompanied by detailed outcome annotations, eventually allowing other researchers to use this benchmark to validate outcome prediction models.
In summary, our work will have a major impact on the implementation of DL in IVF, by providing a much-needed benchmark, ultimately benefiting infertile patients with improved clinical success rates.

\section{Acknowledgments}

The authors would like to thank the IVF staff at the University Hospital of Nantes, and more specifically Dr. Arnaud Reignier and Mrs. Jenna Lammers for the annotation of the database. 
This work was funded by ANR - Next grant DL4IVF (2017)
None of the authors report having competing commercial interests concerning the submitted work.

\bibliographystyle{ieee}
\bibliography{references}

\appendix 

\section{Dataset difficulty.}

To evaluate the difficulty of this dataset, it is not possible to directly compare the baseline performance obtained here with the baseline performance of previous work as they use different sets of classes. 
However, given that previous work use restricted sets of classes compared to the set used in this work, we propose to re-evaluate the baseline models while ignoring and merging classes to obtain a set of classes that is similar to those found in the literature.
More precisely, we first evaluated the baseline ability to identify distinguish early cleavages phases as is often done in the literature \cite{WeakSupMorphoKin,predSuccRate,cellCount,BlastCellCount,ICLRMorpho,DLDP} and secondly we evaluated their ability to discriminate between blastocyst vs not-blastocyst as proposed by \cite{inexpAutoDeepLearn}. 

To reproduce the early cleavage set of classes often used by previous work, we removed test images belonging to phases before p2 and after p9+ and merged classes from p5 to p9+ into one class called p5+.
Using this setup, we obtained similar accuracies to those found in the literature: 0.86, 0.88 and 0.88 for ResNet, ResNet-LSTM and ResNet-3D vs 0.82 to 0.87 in \cite{WeakSupMorphoKin,predSuccRate,cellCount,BlastCellCount,ICLRMorpho,DLDP} (\cref{res_vs_othersetups}). 
To test the performance of blastocyst identification, we processed the predictions made during the first evaluation and merged phases from tPB2 to tM for the non-blastocyst class and merged phases from tB up to the end for the blastocyst class. We ignored the phase pSB as it is a phase of transition to the blastocyst stage not belonging to either of the 2 groups. We obtained accuracies of 0.98, 0.99 and 0.99 vs 0.96 in \cite{inexpAutoDeepLearn} on blastocyst/non-blastocyst evaluation (\cref{res_vs_othersetups}).
Given that baseline performance on this dataset is close to the baseline performance found in previous work, we can conclude that our database in similar in difficulty to previous datasets.

\begin{table}[ht!b]
\makebox[\textwidth][c]{
\begin{tabular}{ccc}\toprule
\multirow{3}{*}{Model}&Identification & Blastocyst\\
     &   of phases       & vs \\
     &      from p2 to p5+                    & Not-blastocyst \\    
\midrule
ResNet&0.86&0.98 \\
ResNet-LSTM&0.88&0.99 \\ 
ResNet-3D& 0.88& 0.99 \\
\bottomrule
\end{tabular}}
\caption{Evaluation of baselines on the identification of phases from p2 to p5+ and blastocyst vs not-blastocyst. The metric used is the accuracy $p$.} 
\label{res_vs_othersetups}
\end{table}

\end{document}